# STOCHASTIC LOTKA-VOLTERRA SYSTEMS OF COMPETING AUTO-CATALYTIC AGENTS LEAD GENERICALLY TO TRUNCATED PARETO POWER WEALTH DISTRIBUTION, TRUNCATED LEVY DISTRIBUTION OF MARKET RETURNS, CLUSTERED VOLATILITY, BOOMS AND CRACHES*

SORIN SOLOMON
*Racah Institute of Physics*
*Givat Ram, Hebrew University of Jerusalem*
*E-mail: sorin@vms.huji.ac.il; http://shum.huji.ac.il/~sorin*

**ABSTRACT**

**We give a microscopic representation of the stock-market in which the microscopic agents are the individual traders and their capital. Their basic dynamics consists in the auto-catalysis of the individual capital and in the global competition/cooperation between the agents mediated by the total wealth invested in the stock (which we identify with the stock-index). We show that such systems lead generically to (truncated) Pareto power-law distribution of the individual wealth. This, in turn, leads to intermittent market (short time) returns parametrized by a (truncated) Levy distribution. We relate the truncation in the Levy distribution to the (truncation in the Pareto Power Law i.e. to the) fact that at each moment no trader can own more than the current total wealth invested in the stock. In the cases where the system is dominated by the largest traders, the dynamics looks similar to noisy low-dimensional chaos. By introducing traders memory and/or feedback between individual and collective wealth fluctuations (the later identified with the stock returns), one obtains clustered volatility as well as market booms and crashes. The basic feedback loop consists in:**
**● computing the market price of the stock as the sum of the individual wealths invested in the stock by the traders and**
**● determining the time variation of the individual trader wealth as his/her previous wealth multiplied by the stock return (i.e. the variation of the stock price).**

The financial markets display some puzzling universal features which while identified long ago are still begging for explanation. Recently, following the availability of computer for data acquisition, processing and simulation these features came under more precise, quantitative study [1]. We first describe phenomenologically these effects and then the unified conceptual framework which accounts for all of them.

## 1   Phenomenological Puzzles

---

*1.1 Pareto Power Distribution of Individual Wealth*

Already 100 years ago, it was observed by Vilifredo Pareto [2] that the number of people *P(w)* possessing the wealth *w* can be parametrized by a power law:

$$P(w) \sim w^{-1-\alpha} \qquad (1)$$

The actual value for $\alpha$ in Eq. (1), was of course only approximately known to Pareto (due to the limited data at his disposal) but, modulo improvement in the accuracy of the statistics it to has been constantly $\alpha=1.4$ throughout the last 100 years in most of the free economies. This wealth distribution Eq. (1) is quite nontrivial in as far as it implies the existence of many individuals with wealth more than 3 orders of magnitude larger than the average. To compare, if such a distribution would hold for personal weight, height or life span it would imply the existence of individual humans 1 Km tall, weighting 70 tons and living 70 000 years.

*1.2 Intermittent market fluctuations with (truncated) Levy-stable distribution of returns*

It was observed by Benoit Mandelbrot [4] in the 60 s that the market returns defined in terms of the time relative variation of the stock index *w* (*t*):

$$r(t,T) = (w(t) - w(t-T))/w(t-T) \qquad (2)$$

are not behaving as a Gaussian (random walk with given size steps) but rather as a so called Levy-stable distribution [3]. In practical terms this means that the probability *Q(r)* for large returns *r* to occur is not proportional to $Q(r) \sim e^{-kr^{**2}}$ as for usual uncorrelated random Gaussian noise but rather to $Q(r) \sim r^{-1-\beta}$. The Levy-stable distribution implies that the actual market fluctuations are very significantly larger than the ones estimated by assuming the usual Gaussian noise. This in turn implies larger risks for the stock market traders and affects the prices of futures, options, insurances and other contracts. The difference between the Gaussian and the Levy-stable distribution is dramatically experienced by traders occasionally, when the gains accumulated over a previous long period of time are lost overnight. The latest market returns measurements [12] show truncation departures from the Levy-stable distribution which we discuss at length in Section 3.

*1.3 Deterministic Chaotic Dynamics of the Global Lotka-Volterra Maps*

With the arrival of the Chaos theories, it was noticed that the prices in the markets behave in a way as if they would be governed by the deterministic motion of a point moving under nonlinear kinematical rules in a space with a small number of dimensions/parameters [5].

One of the earliest chaotic systems was invented in the late 20 s by Lotka and Volterra [6] in order to explain the annual fluctuations in the volume of fish populations in the Adriatic Sea. If one denotes by *w(t)* the number of fish in the year *t*, these discrete logistic equations would predict the fish population *w(t+1)* at the year *t+1* to be:

$$w(t+1) = (1-d+b)w(t) - cw^2(t) \qquad (3)$$

The terms in the bracket in Eq. (3) represent respectively :
- The fish population in the year *t*.

- Minus the proportion (*d*) of fish which die of natural death (or emigrate from the ecology) from one year ( *t* ) to another ( *t+ 1* )
- Plus the proportion (*b*) of fish born (or immigrated) from year *t* to year *t+1*.

The quadratic term $-c\, w^2(t)$ in Eq. (3) originates in the fact that the probability that 2 fish would meet and compete for the same territory /food /mate is proportional to the product $w(t) \times w(t)$. Obviously, only one will obtain the resource and the other one will die, emigrate (be driven out) or be prevented from procreating.

For certain values of the constants *b, d, c,* the population *w*(t) approaches a stationary value: *w(t) = (b-d)/c* but for an entire range of parameters, the system Eq. (3) leads to chaotic annual changes in *w(t)*. To understand it, consider for instance the case in which in the year *t* the population *w(t)* is very small such that the quadratic term $-c\, w(t)^2$ is much smaller than the linear term *(1-d+ b) w(t)*. If *(1-d + b)* is large enough, then the number of fish in the following year *w(t+1) ~ (1-d + b) w(t)* might become quite large. In fact *w(t+1)* may become so large as to make the quadratic term next year $-b\, w^2(t+1)$ comparable to *(1-d+ b) w(t+1)*. The two terms would then cancel and the population *w (t+2)* in the following year would decrease. This may lead to a large *w(t+3)* and so on up and down. It is clear then, that in certain ranges of the parameters, the population will have a quite chaotic (but deterministic) fluctuating dynamics.

However, when people tried to exploit this kind of deterministic systems to predict the markets, it turned out that in fact the fluctuations have in addition to the deterministic chaotic motion very significant random noisy components as well as a continuos time drift in the parameters of the putative deterministic dynamics.

*1.4 Quasi-periodic Market Crashes and Booms*

In addition to the phenomena above, the market seems to have strong positive feed-back mechanisms which in certain conditions reinforce occasional trends and lead to volatility clustering, booms and crashes which we will address later.

## 2  Theoretical Explanations

It came [7]as a surprise, 70 years after the introduction [6] of systems of the type Eq. (3), 100 years after the discovery [2] of the Pareto law Eq. (1) and about 30 years after noticing [4] the intermittent market fluctuations (1.2) that, in fact the generalized discrete logistic (Lotka-Volterra) systems (Eq. (3) and Eq. (5) below), do lead [7] generically to the effects 1.1.-1.3.

Moreover we have constructed realistic market models [8,20] which represent explicitly the individual investors and their market interactions and showed [9] that these realistic models reduce effectively to generalized Lotka -Volterra dynamics Eq. (5) and that these models lead generically to the intrinsic emergence of occasional crashes and booms (effect 1.4). Let us explain how all this happens.

*2.1 Generic Scaling Emergence in Stochastic Systems with Auto-catalytic Elements*

Consider the population **w**(t) of the Lotka-Volterra system Eq. (3) as a sum of *N* sub-populations (families, tribes, sub-species) indexed by an index *i* :

$$w(t) = w_1(t) + w_2(t) + ... + w_N(t) \qquad (4)$$

Obviously, the evolution equations for the $w_i(t+1)$ s have to be such as to reproduce upon summation the discrete logistic Lotka-Volterra system Eq. (3) for $w(t+1)$. We are therefore lead to postulate that the parts $w_i(t)$ fulfill the following dynamics.

At each time interval $t$ one of the $i$ s is chosen randomly to be updated according to the rule:

$$w_i(t+1) = \lambda(t) w_i(t) + a(t) w(t) - c(t) w(t) w_i(t) \qquad (5)$$

The variability originating in the individual local conditions is reflected in the dependence of the coefficients $\lambda(t)$, $a$ and $c$ on time. In particular, $\lambda(t)$ are random numbers of order 1 extracted each time from a strictly positive distribution $\Pi(\lambda)$ while $a$ and $c$ are much smaller than $1/N$. This dynamics describes a system in which the generations are overlapping and individuals are continuously being born (and die) (first term in Eq. (5)), diffusing between the sub-populations (second term) and compete for resources (third term).

*2.1.1 Emergence of Truncated Power Laws Distribution in Stock Invested Wealth*

The dynamics Eq. (5) describes equally well the evolution of the wealth $w_i(t)$ of the individuals $i$ in a financial market. More precisely Eq. (5) represents $w_i(t+1)$ the value of the stock owed by the investor $i$ at time $t+1$ in terms his/her stock wealth $w_i(t)$ at time $t$:

- The first (and most important) term in Eq. (5) expresses the auto-catalytic property of the capital: its contribution to the capital $w_i(t+1)$ at time $t+1$ is equal to the capital $w_i(t)$ at time $t$ multiplied by the random factor $\lambda(t)$ which reflects the **relative** gain/loss which the individual incurred during the last trade period. (This property is consistent with the actual phenomenological data which indicate that the distribution of individual incomes is proportional to the distribution of individual wealth).
- The second term $+ a(t) w(t)$ expresses the auto-catalytic property of wealth at the social level and it represents the wealth the individuals receive as members of the society in subsidies, services and social benefits. This term is proportional to the general social wealth *(/index)* $w(t)$.
- The last term $- c(t) w(t) w_i(t)$ in Eq. (5) originates in the competition between each individual $i$ and the rest of the society. It has the effect of limiting the growth of $w(t)$ to values sustainable for the current conditions and resources. The effects of inflation or proportional taxation are well taken into account by this term.

With these assumptions, it turns out that Eq. (5) leads to a power law distribution of the instantaneous $w_i(t)$ values i.e. the probability $P(w_*)$ for one of the $w_i(t)$ s to take the value $w_*$ is:

$$P(w_*) \sim w_*^{-1-\alpha} \text{ with typically } 1 < \alpha < 2. \qquad (6)$$

Generically, the origin of the distribution $P(w_*)$ in Eq. (6) can be traced to its scaling properties. More precisely, the probability distribution of the form Eq. (6) is the only one whose shape (parameter $\alpha$) does not change at all upon a rescaling transformation

$$w_i(t) \to 2 w_i(t) \qquad (7)$$

In fact, by a rescaling Eq. (7), the Eq (6) transforms into itself with the same $\alpha$:

$$P_\alpha(w) \to P_\alpha(2w) \sim (2w)^{-1-\alpha} \sim w^{-1-\alpha} \sim P_\alpha(w)$$

in contrast to e.g. the Gaussian $Q_k(r)$ whose shape (parameter $k$) changes ($k \to 4k$):

$$Q_k(r) \to Q_k(2r) \sim e^{-k(2r)^{**2}} \sim e^{-4k\,r^{**2}} \sim Q_{4k}(r) \neq Q_k(r).$$

It is therefore expected that such distributions Eq. (6) will be the result of dynamics which is invariant itself under the rescaling Eq. (7). This turns out to be the case of the generalized Lotka-Volterra dynamics generated by Eq. (5). Indeed, if one applies the scaling transformation Eq. (7) on one of the $w_i(t)$ s in Eq. (5) one obtains:

$$2w_i(t+1) = 2\lambda(t)\,w_i(t) + 2a(t)\,w_i(t) - 2c(t)\,u_i(t)\,w_i(t) + a(t)\,u_i(t) - 4c(t)\,w_i^2(t) \qquad (8)$$

where we noted $u_i(t) = w(t) - w_i(t)$.

One sees that $w_i(t+1)$ updated according to Eq. (8) has the same dynamics as before the rescaling transformation (i.e. updating according to the Eq. (5)) except for the last 2 terms in Eq. (8) which are not scaling invariant. Therefore, one expects the scaling power law Eq. (6) to hold for values of the $w_i(t)$ s for which these last 2 terms are negligible in comparison with the other terms in Eq. (8) (recall $a, c << 1/N$):

$$w/N < w_i(t) < w \qquad (9)$$

This is quite a considerable range of $w_i(t)$ s if the system has a large number of elements $i=1,...,N$ (of course for the single variable Lotka Volterra Eq. (3) the scaling range Eq. (9) does not exist as $N=1$ and there is no differentiation between $w_i(t)$ and $w(t)$).

The explicit simulation of Eq. (5) confirms this prediction [7,11,19]: the individual wealth distribution of the $w_i(t)$ s fulfills a power law Eq. (6) **truncated** to the range Eq. (9). The dynamical mechanism is best visualized in terms of the logarithm [9] of the investors relative wealth (normalized to the index $w$): $v_i(t) = \ln(w_i(t)/w(t))$. If one starts with all the traders having the same wealth (delta function distribution at $v_i(0) = -\ln N$), after a relatively short time, the distribution will diffuse into a spreading log-normal distribution. Upon drifting into the lower cut-off induced by the term $\ln a$ (cf. Eq. (5)), the shape of the $v_i$ s distribution $\wp(v)$ will change into a decreasing exponential. In fact if the diffusion coefficient is $\sigma$ and the drift coefficient is $-\mu$, then $\wp(v)$ fulfills the master equation:

$$\partial \wp(v)/\partial t = \sigma\,\partial^2 \wp(v)/\partial v^2 - \mu\,\partial \wp(v)/\partial v$$

which admits the exponential solution $\wp(v)\,dv \sim e^{-v\mu/\sigma}\,dv$. This exponential distribution for the $v_i(t)$ s corresponds to a power law Eq. (6) with exponent $\alpha = \mu/\sigma$ in the original $w_i$ variables [19]:

$$\wp(\ln w)\,d\ln w \sim e^{-(\ln w)\mu/\sigma}\,1/w\,dw = w^{-1-\mu/\sigma}\,dw$$

*2.1.2 The exponent $\alpha$ does not depend on the trends in $w(t)$*

In spite of the fact that the terms non homogenous in $w_i(t)$ in the Eq. (8) are negligible in the scaling range Eq. (9), they do play a crucial role in fixing the boundary conditions for the scaling dynamics. This in turns determine the effective parameters $\sigma$ and $-\mu$ and therefore the value of the scaling exponent $\alpha$. More precisely, upon substituting the instantaneous ideal value $<w(t)>$ of $w(t)$ (the currently expected value of $w(t)$ neglecting fluctuations and relaxation time Eq.(5)- corresponding roughly to the current fundamental financial stock value):

$$<w(t)> = (<\lambda(t)> -1 + N<a(t)>)/<c(t)> \quad (10)$$

into the Eq. (5), one obtains:

$$w_i(t+1) = (\lambda(t) - <\lambda(t)> + 1 - N<a(t)>) w_i(t) + a(t) w(t) \quad (11)$$

The exponent $\alpha$ in Eq. (6) is then fixed[1] by the condition that the distribution $P(w) dw$ is unchanged by the updating Eq. (11). I.e. the flow of traders leaving a certain wealth level $w$ to become poorer or richer equals the total flow of traders arriving at the wealth $w$ from poorer or richer wealth stations $w/(\lambda-<\lambda>+1-Na)$ upon undergoing the transformation Eq. (11):

$$w^{-1-\alpha} dw = \int \Pi(\lambda) [w/(\lambda - <\lambda> + 1 - N a)]^{-1-\alpha} dw/(\lambda - <\lambda> + 1 - N a) d\lambda \quad (12)$$

where $\Pi(\lambda)$ is the distribution of $\lambda$ used in Eq. (5). This leads to the transcendental equation for $\alpha$ [11,15]:

$$\int \Pi(\lambda)(\lambda(t) - <\lambda(t)> + 1 - N<a(t)>)^\alpha d\lambda = 1 \quad (13)$$

which, given $N$, $a$ and the distribution $\Pi(\lambda)$ of $\lambda$ does not depend on $c$ or on an overall shift in $\lambda$ : $\Pi(\lambda) \to \Pi(\lambda-\lambda_0)$. This equation can be either solved numerically for $\alpha$, or approximately by expanding the bracket around 1 in powers of $\lambda(t) - <\lambda(t)> - N<a(t)>$. This leads for $Na << (<\lambda^2> - <\lambda>^2) = \sigma^2 << 1$ to the approximate solution:

$$\alpha \approx 1 + 2Na/[\sigma^2 + (Na)^2] \quad (14)$$

One sees from Eq. (10) that upon changes in the external conditions (resources, predators, etc.), represented by variations in $<c(t)>$, the total population can change by orders of magnitude without affecting the Eq. (13)-(14) which determine the exponent $\alpha$. Therefore the distribution of the individual wealth fulfills the power law Eq. (6) even in non-stationary conditions when $<w(t)>$ varies continuously in time. In fact, even for $c=0$, when an instantaneous ideal value does not exists and the total wealth (/market index) $w(t)$ increases indefinitely, one observes in each particular moment $t$ a very precise power law distribution among the current $w_i(t)$ values, with a stable $\alpha$ exponent. This is confirmed both by simulations and by the observation of experimental data. If $w$ s are town populations or sub-species, a power law distribution which survives large ecological fluctuations in Eq. (10) is predicted.

Note also that the limit $c=0$, $a \to 0$ is not uniform: Eq. (14) gives then $\alpha = 1$ while it is known that in the total absence of the lower bound term $a$, the multiplicative dynamics leads to an ever-flattening log-normal distribution which approaches $\alpha = 0$. This is of practical importance as indeed most of the systems in nature which display power laws have exponents $\alpha$ between 1 and 2. I.e. even an exceedingly small value of $a$ has very significant consequences. To sharpen this intuition let us consider the model in which the term $a(t) w(t)$ is substituted by a reflecting wall [11] which disallows any $w_i(t)$ to assume values lower than $\varepsilon(t) \equiv \omega_0 w(t)/N$:

$$w_i(t) > \varepsilon(t) \equiv \omega_0 w(t)/N \quad (15)$$

where $0 < \omega_0 < 1$.

This corresponds to a social policy where no individual is allowed to slump below the fraction $\omega_0$ of the average wealth $w(t)/N$ (in the typical western economies $\omega_0 \sim$ 1/3). More specifically, during the simulation of the linearized ($a=0, c=0$) Eq. (5) each time that $w_i(t+1)$ comes out less then $\varepsilon(t)$ Eq. (15), its value is actually updated to $\varepsilon(t)$. As proved below, in this model, the exponent $\alpha$ turns out totally independent on $\Pi(\lambda)$ and for a wide range is determined only by $\omega_0$ through the (approximate) formula [11]:

$$\alpha = 1/(1-\omega_0). \tag{16}$$

---
[1] neglecting in the region Eq. (9) the term $a(t)w(t)$.

Note again that the limit of vanishing lower bound $\omega_0 \to 0$ leads to $\alpha=1$ while for finite $N$ in the absence of a lower bound ($\omega_0 = 0$) the distribution approaches an infinitely flat log-normal distribution corresponding to $\alpha=0$. However, the value $\alpha=1$ is relevant for firms sizes, towns populations and words frequencies where a very small but finite lower bound is provided by the natural discretization of $w_i$ (number of firm employees, town residents, word occurrences cannot be lower than 1).

The Eq. (16) can be obtained starting form the expressions for the total probability and the total wealth. First, using the fact that the total probability equals 1, one fixes (assuming $N/\omega_0 \gg 1$) the constant of proportionality in Eq. (6):

$$1 = C\int_\varepsilon x^{-1-\alpha} dx \Rightarrow 1 \approx C\varepsilon^{-\alpha}/\alpha \Rightarrow C \approx \alpha[\omega_0 w/N]^\alpha \tag{17}$$

Then, using this expression for $C$ in the formula of the total wealth $w$ one gets:

$$w = NC\int_\varepsilon x^{-\alpha} dx \Rightarrow w \approx NC\varepsilon^{1-\alpha}/(\alpha-1) = \alpha\omega_0 w/(\alpha-1) \tag{18}$$

which means (dividing by $w$):

$$1 \approx \alpha\omega_0/(\alpha-1) \tag{19}$$

which by solving with respect to $\alpha$ leads to (16). For very low values of $\omega_0 \ll 1/N$ the upper bound $w$ in the integral in (18) becomes relevant and the value of $\alpha$ does depend on $N$:

$$\alpha \approx -\ln N/(\ln\omega_0 - \ln N) \tag{20}$$

which in particular has the correct limit $\alpha=0$ for $\omega_0 = 0$. For intermediate values, one can solve numerically the couple of equations (17) and (18) with the correct upper bounds ($w$ cf. the truncation Eq. (9)) in the integrals and the result agrees with the actual simulations of the linearized random system Eq. (5) submitted to the lower bound constraint (15). This concludes the deduction of the truncated Pareto power law 1.1 in generic markets (and in general systems consisting of auto-catalyzing parts, ecologies, etc.).

*2.2 Pareto Wealth Distribution implies Levy-Flights Returns*

Let us now understand the emergence of the property 1.2: the fluctuations of the $w(t)$ around its ideal (financially fundamental) instantaneous value $<w(t)>$ Eq. 10 are not Gaussian. The Gaussian distribution is intuitively visualized by imagining a person (drunkard) taking at each time $t$ a step (of approximate unit size) to the left or to the right with equal probability. The Gaussian probability distribution $Q(r)$ is then approximately defined by the probability for the person (drunkard) to end-up after $T$ steps, at the distance $r(T)$ from the starting point. For instance, the probability for the largest fluctuations (say $r(T) = T$) to occur is easy to compute: it is the probability that all the $T$ steps will be in the same direction. This equals the product of the probabilities for each of the $T$ steps separately to be in that direction i.e. ( )$^T$.

At the first sight, this Gaussian random walk description fits well the time evolution of the sum $w$ Eq. (4) under the dynamics Eq. (5): at each time step one of the $w_i(t)$'s is updated to a new value $w_i(t+1)$ and the sum $w(t)$ changes by the random quantity $w_i(t+1) - w_i(t)$. Consequently, the dynamics of $w(t)$ consists of a sequence of random steps of magnitude $w_i(t+1) - w_i(t)$. The crucial caveat is that according to the Eq. (5), for the range Eq. (9) in which the last 2 terms in Eq. (7) are negligible, the steps $w_i(t+1)-w_i(t)$ are proportional to the $w_i(t)$'s themselves. Consequently, $w(t)$ evolves by random steps whose magnitude is **not** approximately equal to any given unit length. Rather, the random steps $w_i(t+1) - w_i(t)$ have various magnitudes with probabilities distributed by a (truncated Eq. (9)) power law probability distribution Eq. (6).

Random walks with steps whose sizes are not of a given scale [3] but are distributed by a (truncated) power law probability distribution Eq. (6) $P(w.) \sim w.^{-1-\alpha}$ are called (truncated) Levy flights [3]. The associated probability distribution $Q_\alpha(r)$ that after $T$ such steps the distance from an initial value $w(t)$ will be $r = (w(t+T) - w(t))/w(t)$ is called a (truncated) Levy-stable[2] distribution of index $\alpha$.

The Levy-stable distribution[3] has very different properties from a Gaussian distribution. For instance the probability that after $T$ steps, the walker will be at a distance $r = T$ from the starting point is this time dominated by the probability that one of the steps is of size $T$. This will happen (according to Eq. (6)) with probability of order $T^{-1-\alpha}$. To see the dramatic difference between the time fluctuations $r = (w(t+T) - w(t))/w(t)$ predicted by the Levy flights vs. the ones predicted by the Gaussian, consider the typical numerical example $\alpha = 1.5$ and $T=25$. The probability that after a sequence of 25 steps, a Levy flyer will be at a distance 25 from the starting point is $25^{-2.5}$ i.e. about 1000 times larger than the corresponding Gaussian walk probability estimated above: $1/2^{25}$.

Therefore, our model Eq. (5), predicts the emergence with significant probability of very large intermittent truncated-Levy distributed market fluctuations.

**In fact we reach a more general conclusion: ANY quantity which is a sum of random increments proportional to the wealths $w_i(t)$ will have fluctuations described by a Levy distribution of index $\beta$ equal to the exponent $\alpha$ of the wealth power distribution Eq. (6) of the $w_i$'s.**

In particular, since the individual investments are stochastically proportional to the investor's wealth (which is consistent with the empirical fact that the income distribution is proportional to the wealth distribution) one predicts [11] that the stock market fluctuations will be described (effect 1.2) by a truncated-Levy distribution of index equal to the measured exponent $\alpha = 1.4$ of the Pareto wealth distribution Eq. (6) (effect 1.1). This highly nontrivial relation between the wealth distribution and the market fluctuations is confirmed by the comparison of the latest available experimental data [12].

In principle alternative mechanisms exploiting the Lotka-Volterra mechanism could take place: the investors could be roughly of the same wealth but act in herds [13-16] of magnitude $w_i$ which evolve according the generalized Lotka-Volterra Eq. (5). This would lead to power law distribution in the size of these herds and consequently to a Levy distribution in the market fluctuations. However unless all the investors in a herd make their bid at the same time, the herding will show in time correlations of the stock evolution. Presently, there is no evidence for such correlations nor for the presence of scaling herds. Still, one should keep an open mind [16,22] about the significance of the $w_i(t)$ s involved in the Eq. (5).

---

[2] The name "stable" has to do with another property of these distributions which is not directly relevant to this report.

[3] The cut-off in the Levy-stable distribution for large values of $r$ is discussed in Section 3 below.

*2.3 Levy-Stable fluctuations may look like noisy, drifting, deterministic Chaos*

In usual stochastic systems in which the elementary degrees of freedom $w_i(t)$ are of the same order of magnitude, the fluctuations around the mean value are the result of a large number of random contributions of the same size: Gaussian noise.

In the case in which the elementary degrees of freedom are distributed according to Eq. (6) the largest degrees of freedom are macroscopic and, in certain cases (especially for low $\alpha$), the dynamics is dominated by the few largest elements.

This might look locally as a chaotic process of low dimensionality (small number of parameters and of degrees of freedom). However, the relevant $w_i(t)$ s may change in time and the smaller $w_i(t)$ s do imply additional fluctuations not related with the dominating degrees of freedom. Therefore, while in certain limits one may obtain deterministic low dimensional chaos, this is only an extreme idealization.

*2.4 LLS model: Market Crashes; Strategies Evolution; How Generic is Scaling ?*

The model Eq. (5) was initially proposed as a mezoscopic description of a wide series of simulation experiments on the (Levy-Levy-Solomon [8,20-22]) **LLS** microscopic representation model. The LLS model considers [8] individual investors with various procedures of deciding the amount of stock to sell/buy at each time and studies the resulting market dynamics. The variants of the basic model displayed always the central features of Eq. (5). In particular, for a very wide range of wealth Eq. (9) the gain/loss of each investor was stochastically proportional to its current wealth. So was its influence on the market changes. Therefore the ingredients necessary to obtain the effects 1.1.-1.3 through the mechanisms described above were always in place in the realistic LLS market models.

Moreover, a very simple mechanism for booms and crashes (effect 1.4) was naturally present in many of the LLS runs [8]: each investor had the tendency to assume that the future behavior of the market is going to be similar to the past one. This meant that prices had the tendency to rise beyond the value justified by the expected dividends (the investors expected gain from the sheer increase in the stock price). This lead to a quite unstable situation as the discrepancy between the dividends and the market price level became more and more severe. At a certain stage, an usual, relatively mild downwards fluctuation would take place. As this downwards fluctuation was internalized as part of the traders past experience, it lead to a lowering of the future expectations, followed by a further decrease in prices. This iterated and triggered eventually an avalanche effect. The down trend would stop only as the market price became so low that the expectation of the dividends alone justified buying the stock. Therefore, even the simplest versions of the LLS model are capable to generate in addition to the universal features 1.1-1.3 also the cycles of crashes and booms characteristic for the real markets.

The simple LLS model was also capable to create interesting dynamics in the strategies of the investors. Indeed, as it happened often [20], the investors having a particular investing policy would be occasionally advantaged by the current dynamics of the market. This would cause them to earn more from the market fluctuations. As they became richer, they influenced increasingly the market changing thereby its dynamics. In the new dynamical state, another group would become advantaged and start to become richer thereby changing in its turn the character of the market. The new market behavior would be advantageous for yet another group and so on. Therefore the market would pass through various epochs without external influence and without any of the investors changing its strategy [20].

Recently, models considering strategy changes have been proposed [5,16,22], which may lead to yet richer results. The inclusion of adaptive agents, transaction costs, herding effects etc. in the LSS model is quite straightforward but may lead to further uncovering of unexpected and highly non-trivial relationships and results.

Finally, one can consider the possible extensions of the dynamics Eq. (5) to even more generic systems. First, one can use instead of the average *w(t)* Eq. (4) any combination of the $w_i(t)$ s with positive coefficients or a general function *W* $(w_1, ... w_N)$ which depends monotonically (increasingly) in each of the $w_i(t)$ s. Secondly the random distribution of the $\lambda$ s, and the constants *a* and *c* can be included in a more general form [7]:

$$w_i(t+1) = \Phi(t,W) \, w_i(t) + \Psi(t,W) \qquad (21)$$

where $\Phi$ and $\Psi$ are such that there exists an equilibrium region of *W* s for which the function $<\Phi(W)> - 1$ changes sign and for which $<\Psi(W)>$ is much smaller than 1. Assuming the $w_i$ s very small compared with *W*, one can consider the generic stochastic Arecchi system:

$$w_i(t+1) - w_i(t) = \partial \Lambda(t,W) \, / \, \partial w_i \qquad (22)$$

where $\Lambda$ is a random potential whose partial derivatives admit Eq. (21) as an approximate linear expansion in the $w_i$ s around its (dynamical) ground state. The properties 1.1-1.3 will then still hold for a wide range of parameters.

## 3  Corrections to Levy-flight behavior. Relation to (G)ARCH models

Let us now return to the dynamics of the system Eq. (5) and study the departures from a perfect uncorrelated Levy-flight behavior. This is of more than academic interest in as far as real markets display clearly the existence of such departures both in the market returns distribution and in the time correlations (especially of the market volatility ).

*3.1 Truncated Pareto Power Laws lead to Truncated Levy-Flights*

Let us first discuss the departures from the ideal Levy-stable shape of the market fluctuations (returns) assuming for the moment that the individual steps (cf. Eq. (11)) in the range Eq. (9))

$$w_i(t+1) - w_i(t) \sim K \, w_i(t) \equiv (\lambda - <\lambda> - Na) \, w_i(t)$$

are independent.
In order to obtain a perfect Levy-stable shape it is necessary that the sizes of the individual steps of the Levy-flights are distributed perfectly by the power law Eq. (6) up to arbitrary large steps sizes. This is obviously inconsistent with the wealth truncation Eq. (9) which implies that steps of size K $w_i(t) > K_{max} \, w(t)$ have 0 probability to appear. Therefore, steps of arbitrary size expected from the Levy-flight model fail to appear in the model Eq. (5): the Pareto Power-law in $w_i$ Eq. (6) is truncated and consequently the corresponding Levy-stable distribution in *r (t)* Eq. (2) is truncated. This starts to be quantitatively relevant for times *T* for which according to the untruncated distribution, the probability of elements $w_i(t)$ of size $w$ to appear is of order 1. This $T$ can be estimated by estimating the probability (in the absence of truncation) for a $w_i$ to equal or exceed the entire wealth $w$ in the system (cf. Eq. (17)):

$$P(w) = C w^{-\alpha} / \alpha \approx w^{-\alpha}[\omega_0 w / N]^{\alpha} = [\omega_0 / N]^{\alpha} \qquad (23)$$

Therefore, in a non-truncated Levy flight, one would expect for times $T \geq [N/\omega_0]^{\alpha}$ the appearance, with high probability, of $w_i(t)$ s of order $w(t)$ and larger. Since such $w_i(t)$ s will fail to show up, the distribution looks truncated. For $T$ s of the order $T \sim [N/\omega_0]^{\alpha}$, the truncation will just consist in the probability being 0 for fluctuations beyond the magnitude $K_{max} w(t)$. For even larger times, the truncation will appear trough the fact that the largest fluctuations will be dominated by the probability to have a few individual steps of the largest allowed size $K_{max} w(t)$ rather than one single step of larger (un-allowed by Eq. (9)) magnitude. This will lead to an exponential tail rather than the naively expected untruncated Levy-stable law. For times $T > N^{1+\alpha}$ this Gaussian mechanism replaces the Levy-flight dynamics over the entire distribution.

These departures from the Levy-stable shape are actually compounded with deviations from the assumption of stochastic independence of the successive $w(t)$ increments $w_i(t+1) - w_i(t)$. Indeed, the value of the $w_i(t+1)$ depends through the linear term $a$ and especially through the bilinear term $-cw_i(t)w(t)$ on the other $w_i(t)$ s at previous times. In particular, if the updating of a large $w_i(t)$ induces a large departure of $w(t)$ from its ideal value $< w(t) >$, this departure will survive for a while and it will influence the dynamics of the updatings of the subsequent $w_i(t)$ s. Therefore, one expects that the dynamics Eq. (5) will lead to time correlations in the fluctuations amplitudes. This kind of effects are recorded in actual market measurements and it is desirable to take advantage of the various possible mechanisms to generate and control them.

*3.2 Total stock wealth closes feedback loop: investors wealth $\Leftrightarrow$ market returns*

A particular, a very important extension of (5) is to make the distribution $\Pi(t,\lambda)$ of the $\lambda$ s in Eq.(5) dependent on the $w_i$ s or of their time variations. For instance, one can dynamically update the distribution $\Pi(t,\lambda)$ to equal the distribution of the returns $r$ during the time period preceding $t$. This would induce time correlations in the volatility and clustering similar to the ones observed in nature and mimed by the various ARCH models [23]. The difference would be that, in our case, the distribution will also reproduce more realistically the (truncated) Levy character of the fluctuations.

Another dependence of the $\lambda$ s on the previous dynamics is suggested by the intuition that the actual increase or decrease $K w_i(t)$ in the capital invested in stock by an individual trader $i$ is a result of the actual trends (returns $r$) in the market in the time passed since his/her last trade(s). This can take place both because the direct influence of the market on the trader s wealth as well as through the market trends influence on his/her investment decisions. Limiting oneself only to the influence of the return since the last trade (this can be relaxed by taking into account longer memories [8,20]) one is lead to the dynamics:

$$w_i(t+1) = (\eta(t) + \mu(t) r_i(t)) w_i(t) + a(t) w(t) - c(t) w(t) w_i(t) \qquad (24)$$

where $\mu$ and $\eta$ are parameters and the return $r_i(t)$ since the last trade is defined by:

$$r_i(t) = (w(t) - w(t_i))/w(t_i) \qquad (25)$$

where the time $t_i$ in Eq. (25) is the last time the element $i$ was updated. Recall that at each time $t$, there is one single element $i$ which is randomly selected for being updated by the Eq. (5) respectively Eq. (24)). Therefore the term $w(t)-w(t_i)$ couples the past fluctuations in the market to the individual capital fluctuations. The random character of the terms $\mu(t)$ and $\eta(t)$ is not mandatory. In its absence, the random selection of the updated element $i$ is the only source of randomness in (24). The new multiplicative term $r_i(t)$ in Eq. (24) introduces positive feedback in the dynamics of the $w_i$ s: an increase in $w_i$ leads to an increase in $w$, which leads to an increase in $r$ which leads to an increase in $w_i$ closing the positive feedback loop. In certain cases this can lead to crashes in other macroscopic events. Following the dynamics Eq. (24), both $r_i$ and $w_i$ will acquire long power-like tails and one expects a 2-powers tail in the market price $w(t)$ fluctuations of the type lately measured in real markets [13].

A variety of sub-models can be constructed with personalized ($i$-dependent) investment strategies $X_i(t, w_i(.))$ which are (possibly stochastic) functionals of the past stock market **history** $w_i(.)$. In particular the $X_i(t, w_i(.))$ s may simulate adaptive technical trading strategies and/or portfolio optimization (by investing in more than one instrument $K$: $w_i^K(.)$ ). Moreover, one may introduce an active monetary policy by considering adaptive and history ($w_j^M(.)$) dependent subsidies $E(w_j^M(.))$ and taxes/trading costs $F(w_j^L(.)) w_i^K(t)$:

$$w_i^K(t+1) = X_i^K(t, w_j^L(.)) [(1+ r_i^M(t)) w_i^M(t) + E(w_j^L(.))] - F(w_j^L(.)) w_i^K(t) \quad (26)$$

Moreover, rather than assuming equal trading probability for all the investors, the very probability for the element $i$ to trade at time $t$ may be treated as an adaptive $i$-dependent functional $\pi_i(t, w_j^M(.))$ depending on the market history $w_j^M(.)$.

The crucial new idea is:
- the identification of the total wealth $w^M(t)$ invested by the traders in a stock M:

$$w^M(t) = \sum_i w_i^M(t) \quad (27)$$

as an index of the market price of the stock M and
- the use of the resulting M-stock returns:

$$r_i^M(t) = (w^M(t) - w^M(t_i))/w^M(t_i) \quad (28)$$

(where $t_i$ is the last time the investor $i$ traded) in the updating of the current wealth $w_i(t)$ of the individual $i$:

$$w_i(t+1) = \sum_M (1+ r_i^M(t)) w_i^M(t) \quad (29)$$

Adaptive investor strategies $X_i^K(t, w_j^M(.))$, taxes and trading costs $E(w_j^M(.))$, $F(w_j^M(.))$, herding effects, labor and external market effects etc. [14,16,18,22,24] can then be introduced cf. Eq. (26).

The Unstable-Stauffer system (**USS**) Eq. (26)-(29) is an efficient compromise between the microscopic representation LLS model [8,20], the mezoscopic generalized Lotka-Volterra model Eq. (5) and the macroscopic models of the ARCH type [23].

## 4    Acknowledgments

This research was supported in part by the USA-Israel Binational Science Foundation, the Germany-Israel Foundation. I thank the co-authors of my papers quoted in this report: O. Biham, H. Levy, M. Levy, O. Malcai, N. Persky. Dietrich Stauffer provided much of the energy for this quite long series of papers. Tito Arecchi suggested the dynamics of the type Eq. (22) for the study of chaotic systems.

# 5 References

1) S. Solomon, in *Annual Reviews of Computational Physics* I, Ed. D. Stauffer, p. 243 (World Scientific, 1995).
2) Pareto, V. Cours d Economique Politique}, Vol 2, (1897).} *Manual of Political Economy*, Macmilan. G.K. Zipf, *Human Behaviour and the Principle of Least Effort* (Addison Wesley Press, Cambridge MA 1949).
3) P. Levy, *Theorie de l Addition des Variables Aleatoires* Gauthier-Villiers, Paris, 1937 . E.W. Montroll and M.F. Shlesinger, Proc. Nat. Acad. Sci. USA 79 (1982)3380. R.N. Mantegna and H.E. Stanley in *Levy Flights and Related Topics in Physics*, Eds. M.E. Shlesinger, G. Zaslavsky and U. Frisch, Springer 1995.
4) Mandelbrot, B. B. J. Business 36, 394-419 (1963); *The Fractal Geometry of Nature*, Freeman, San Francisco, 1982), Econometrica 29 (1961) 517.
5) P.W. Anderson, J. Arrow and D. Pines, eds. *The Economy as an Evolving Complex System*, Redwood City, Calif.: Addison-Wesley,1988
6) J. Lotka, *Elements of Physical Biology*, Williams and Wilkins, Baltimore 1925. V. Volterra, Nature 118(1926)558; R.M. May, Nature 261(1976)207.
7) S. Solomon and M. Levy International Journal of Modern Physics C , Vol. 7, No.5 (1996) 745; O. Biham, M. Levy, O. Malcai and S. Solomon preprint.
8) M. Levy, H. Levy and S. Solomon, Econ. Lett. 45(1994)103; J.Phys 5(1995)1087.
9) H.Simon, Biometrika, vol42, Dec.1955;Y. Ijiri and H.Simon, *Skew Distributions and the Sizes of Business Firms*,(North Holland, Amsterdam, 1977).
10) Gell-Mann, M. ,*The Quark and the Jaguar*, Little Brown and Co.,London 1994.
11) M. Levy and S. Solomon Int. J. Mod. Phys. C , Vol. 7 (1996) 595;
12) R. N. Mantegna and H. E. Stanley, Nature 383 (1996) 587; 376 (1995) 46.
13) Potters, M., Cont, R. and Bouchaud, J.P. Europhys. Lett 41(1998)3
14) Bak, P; Paczuski, M; Shubik, M. Physica A 3-4 (1997) 430. R. Cont and J-P. Bouchaud, cond-mat 9712318.
15) D. Sornette and R. Cont J. Phys. I France 7, (1997) 431; Kesten, Acta. Math. 131 (1973) 207 .
16) M Marsili, S. Maslov and Y-C Zhang, cond-mat/ 9801239, Caldarelli, G; Marsili, M; Zhang, YC., Europhysics Letters 40 (1997) 479, Challet, D; Zhang, YC., Physica A 246 (1997) 407
17) M. Levy and S. Solomon Physica A 242 (1997) 90.
18) Takayasu et al., Physica A 184 (1992) 127.
19) M. Levy and S. Solomon, Int. J. Mod. Phys. C, Vol. 7 (1996)65.
20) M. Levy, N. Persky and S. Solomon Int. J. of High Speed Computing 8 (1996)93
21) T. Hellthaler, Int. J. Mod. Phys. C 6 (1995) 845; D. Kohl, Int. J. Mod. Phys. C.
22) S. Moss De Oliveira, P.M.C de Oliveira, D. Stauffer, *Sex, Money, War and Computers: Nontraditional Applications of Computational Statistical Mechanics*, in press 1998.
23) BaillieR. and Bollerslev, T. Rev. of Econ. Studies 58 (1990) 565; Engle R.F., Ito T. and Lin W.-L. Econometrica 58 (1990) 525.
24) R.G. Palmer, et. al. , Physica D, 75, (1994) 264.